
-------------------------------------------------------------------------------
%
%
\input phyzzx
\baselineskip=24pt

\def\smallspace{\baselineskip=12pt}
\def\bigspace{\baselineskip=24pt}

\def\sscr{\scriptscriptstyle}

%
%
\nopagenumbers
\hsize=5.2truein
\hoffset=0.65truein
\smallspace
\line{\hfill\vtop{\hbox{IP-ASTP-03-93}\hbox{Feb. 1993}}}
\bigspace
\vskip 1.35truein
\centerline{\bf Cosmological Bound on the Decay $\pi^0\rightarrow\gamma X$}
\vskip 1.0truein
\centerline{Kin-Wang Ng}
\vskip 0.2truein
\baselineskip=16pt
\centerline{\it Institute of Physics, Academia Sinica,}
\centerline{\it Taipei, Taiwan 11529, Republic of China}
\vskip 1.7truein
\centerline{\bf Abstract}

Using the upper bound on the effective number of light neutrino species during
primordial nucleosynthesis and the cosmological pion-pole mechanism
$\gamma\gamma\rightarrow \pi^0\rightarrow \gamma X$,
we obtain an upper limit on the branching ratio for the decay
BR$(\pi^0\rightarrow \gamma X)<3\times 10^{-13}$, where $X$ is any long-lived
weakly interacting neutral vector particle with mass smaller than the neutral
pion mass.
\vfill\eject
%
%
\footline={\hss\tenrm\folio\hss}
\hsize=6.5truein
\hoffset=0.0truein
\bigspace
Experimental searches for exotic neutral pion decays have been reported
recently [1,2]. These experiments serve not only to test the Standard Model,
but also to hunt for new physics. So far, the experimental results
are consistent with standard physics. As such, upper limits on the
branching ratios for the rare decays
BR$(\pi^0\rightarrow\nu {\bar\nu})<8.3\times 10^{-7}$ [1] and
BR$(\pi^0\rightarrow\gamma X)<5\times 10^{-4}$ [2] have been set,
where $X$ is a hypothetical long-lived weakly interacting neutral particle
with mass $0\le m_x < m_{\pi^0}$. From the conservation of angular momentum,
the
decay $\pi^0\rightarrow\nu{\bar\nu}$ is allowed only if the right-handed
(RH) neutrino
exists. Also, observation of such a radiative decay of pion would indicate the
unambiguous existence of a new vector particle, which might be a new
$U(1)$ gauged boson proposed in some extended standard models [3,4,5].

On the other hand, recently, we have considered the cosmological bound on the
decay $\pi^0\rightarrow\nu{\bar\nu}$ [6]. If RH neutrinos exist, they
would be produced from the cosmic thermal background via the
pion-pole mechanism ($\pi$PM), $\gamma\gamma\rightarrow \pi^0\rightarrow
\nu{\bar\nu}$, at the temperature of about the pion rest mass [7]. As long as
the neutrino lives longer than the time scale for the freeze-out of weak
interactions ($\simeq$ 1 sec),
in order that the amount of RH neutrinos thus produced would not exceed the
primordial nucleosynthesis bound on the effective number of light neutrino
species, $N_\nu\le 3.3$ [8], the branching ratio for the decay
BR$(\pi^0\rightarrow\nu {\bar\nu})$ must be less than $3\times 10^{-13}$ [6],
which is much better than the experimental limit. Here, in a similar way, by
calculating the amount of $X$ bosons produced from the cosmic photons via the
$\pi$PM, we will set, from the nucleosynthesis bound, an upper bound on the
decay $\pi^0\rightarrow \gamma X$.

It was Fischbach et al. [7] who first applied the $\pi$PM
to astrophysics and cosmology.  The cross section for the reaction $\gamma
\gamma\rightarrow\gamma X$ via a pion resonance can be approximated by
$$
\sigma(s) = {{ 8\pi\; \Sigma(s) \;\Gamma(\pi^0\rightarrow
\gamma\gamma) \;\Gamma(\pi^0\rightarrow \gamma X)} \over
{(s-m^2_\pi)^2 + m_\pi^2 \Gamma_\pi^2} }\;,\eqno(1)$$
where $m_\pi\simeq 135$ MeV and $\Gamma_\pi\simeq 7.83$ eV are, respectively,
the mass and width
of pion, and $\sqrt{s}$ is the center-of-mass energy; $\Sigma(s)$ is
generally a polynomial of $s$ with $\Sigma(s=m_\pi^2)=1$, which would take the
functional form,
$$
\Sigma(s)=\left( {{s-m_x^2}\over {m_\pi^2-m_x^2}}\right)^3 \;,\eqno(2)$$
if $X$ is a vector boson directly coupled to quarks or leptons [3]. Because
$\Gamma_\pi/m_\pi \simeq 5.8\times 10^{-8}$ is quite small, we would have
$\Sigma(s) \simeq 1$ at $s$ near the pion pole (when $s=m_\pi^2$)
unless $m_x$ is very close to $m_\pi$.
Thus, in general, the cross section $\sigma(s)$ is much smaller at $s$ far
from the pion pole than at $s$ near the pole. Note that Eq. (1)
gives the correct off-shell scattering cross section at large $s$.
If the photons are in thermal equilibrium with a temperature $T$, the
production rate of $X$ bosons is given by
$$
Q={1\over 2}\;{\langle \sigma v_{\sscr rel}\rangle}\; n_\gamma^2 \;,\eqno(3)$$
where $\sigma$ is the cross section given by Eq. (1), $v_{\sscr rel}$ is the
relative velocity of the incident photons, and $n_\gamma = [2\zeta(3)/\pi^2]\;
T^3 \equiv AT^3$ ($\zeta(3)=1.202$) is the
number density of the thermal photons. In Eq. (3), the brackets imply a thermal
average and the factor of $1\over 2$ appears due to identical initial photons.
At temperature $T$, the average energy per photon is
${\langle E_\gamma \rangle} = [\pi^4/30\zeta(3)]\;T \equiv CT$.
In the following, we shall use the approximation,
${\langle \sigma v_{\sscr rel}\rangle} \simeq \sigma(s)$, where
$s=4\;{\langle E_\gamma \rangle}^2$.

Now we turn to the calculation of the amount of $X$ bosons produced from
the cosmic thermal background.  Since the whole process takes place in a
radiation-dominated universe, the number density $n_x$ of $X$ bosons
at time $t$ is governed by the Boltzmann equation,
$$
{{dn_x}\over {dt}} = - 3Hn_x + Q\;,\eqno(4)$$
where $Q$ is given by Eq. (3) and $H$ is the Hubble parameter given by
$$
H = \left[ {{8\pi G}\over 3}\; {{\pi^2}\over {30}}\; g_*(T)\right]^{{1\over 2}}
T^2 \equiv BT^2\; . \eqno(5)$$
Here, $G$ is Newton's constant, and $g_*(T)$ is the total number of
degrees of freedom at temperature $T$.  In Eq. (4), we have neglected the
inverse scattering processes.
Since $R\propto 1/T$, where $R$ is the cosmic scale factor, by writing
$n_x=f(T)\;T^3$, we obtain from Eqs. (1), (3) and (4) that
$$
{df\over {dT}} = -{{A^2}\over {2B}}\;\sigma(s)\;,\eqno(6)$$
where $s=4C^2 T^2$.
Eq. (6) can be directly integrated and we obtain
$$
f(T_1)-f(T_0)={{2\pi A^2}\over {BC}}\;{{\Gamma_\pi^2 R}\over {m_\pi^3}}
\int^{\ \ \tau_0}_{\tau_1} {{\Sigma(\tau)}\over {(\tau^2 -1)^2 + \gamma^2}}
\;d\tau\;,\eqno(7)$$
where $\tau \equiv \sqrt{s}/m_\pi$, $\gamma \equiv \Gamma_\pi/m_\pi$,
$R \equiv
\Gamma (\pi^0\rightarrow \gamma X)/ \Gamma (\pi^0\rightarrow \gamma\gamma)$,
and we have taken $\Gamma (\pi^0\rightarrow \gamma\gamma) \simeq \Gamma_\pi$.
In Eq. (7),
$T_1\simeq 1$ MeV is the temperature at which weak interactions freeze out.
Note that the upper limit $T_0$ of the integral cannot be arbitarily large
(the integral is divergent at large $T_0$ when $\Sigma(s)$ takes the form
in Eq. (2)). Here, we choose $T_0\simeq 100$ MeV and assume $f(T_0)=0$, based
on the fact
that the $X$ boson has been decoupled from the thermal background at a much
higher temperature and then the number density $n_x$ might be negligibly
small right after the quark-hadron phase
transition which occurs at a temperature of about 100 MeV. Since
$\Sigma(\tau)\simeq 1$ when $\tau \simeq 1$, we can estimate
the integral in Eq. (7) by
$$
\int^{\ \ \tau_0}_{\tau_1} {{\Sigma(\tau)}\over {(\tau^2 -1)^2 +
\gamma^2}}\;d\tau \simeq {k\over \gamma}\;,\eqno(8)$$
where $k$ is a number of the order of 1. By using $\Sigma(\tau)$ as
given in Eq. (2), we find numerically that $k\simeq 1.57$. In fact,
the result is rather insensitive to different functional forms of
$\Sigma(\tau)$, and also to both integration limits
as long as the integration range has covered the pion pole reasonably well.
Since the production of $X$ bosons via the $\pi$PM proceeds at a
temperature $T\simeq 0.185\; m_\pi \simeq 25$ MeV, we take $g_*(T)\simeq
10.75$ and note that there would be no further heating of the thermal bath
due to particle annihilations or phase transitions from $T\simeq 25$~MeV
down to $T\simeq 1$~MeV. Finally, we obtain from Eqs. (7) and (8) that
$$
f(T=1{\rm MeV})=1.57\; {{2\pi A^2}\over {BC}}\; {{\Gamma_\pi R}\over {m_\pi^2}}
\;,\eqno(9)$$
provided that the lifetime of $X$ boson is longer than 1 sec.

So far, we have not restricted the $X$ boson mass $m_x$ except $0\le
m_x<m_\pi$. For $m_x<<1$ MeV, at $T\ge 1$ MeV, the $X$ bosons are relativistic
and their average energy per particle is $\langle E_x \rangle =CT$.
Thus, their energy density $\rho_x$ is given by
$$
\rho_x = n_x \langle E_x \rangle = f(T)\;T^3 CT = 1.57\; {{2\pi A^2}\over
{B}}\; {{\Gamma_\pi R}\over {m_\pi^2}}\; T^4\;.\eqno(10)$$
Expressing $\rho_x$ in terms of $n_{\sscr eff}$, the equivalent number of light
neutrino species, i.e., $\rho_x ={7\over 8}(\pi^2/15)\;n_{\sscr eff}T^4$, we
find $$
n_{\sscr eff}= 9.8\times 10^{11} R\;.\eqno(11)$$
Therefore, in order that the contribution of the
energy density of $X$ bosons to the thermal background during the
time of nucleosynthesis does not exceed the equivalent of 0.3 neutrino
species, i.e., $n_{\sscr eff} < 0.3$, the branching ratio $R$ in Eq. (11)
must be such that
$$
R < 3\times 10^{-13}\;.\eqno(12)$$
For $m_x>>1$ MeV, at $T\ge 1$ MeV, the $X$ bosons are non-relativistic and
$\rho_x$ is simply given by
$$
\rho_x=n_x m_x=f(T)\;T^3 m_x=1.57\; {{2\pi A^2}\over {BC}}\; {{m_x \Gamma_\pi
R}\over {m_\pi^2}}\; T^3\;.\eqno(13)$$
Again, expressing $\rho_x$ in terms of $n_{\sscr eff}$, we find
$$
n_{\sscr eff}=4.9\times 10^{13}{m_x \over m_\pi}\;R \left({{\rm MeV}\over
T}\right)\;.\eqno(14)$$
When $m_x \le m_\pi$, $n_{\sscr eff} < 0.3$ implies
$$
R < 6\times 10^{-15}\;.\eqno(15)$$

In conclusion, we have applied the $\pi$PM to set a cosmological upper
limit on the branching ratio for the decay of a neutral pion into a gamma plus
a weakly interacting neutral vector particle $X$ with mass smaller than the
pion mass and lifetime longer than
1 sec., BR$(\pi^0\rightarrow \gamma X)< 6\times 10^{-15}-\;3\times 10^{-13}$,
depending on the mass of $X$.
This limit is at least nine orders of magnitude better than the experimental
limit.

\vskip 0.5truein
\centerline {\bf Acknowledgments}
This work was supported in part by the R.O.C. NSC Grant No.
NSC82-0208-M-001-131-T.
\vfill\eject
%
%
%
%
\def\et{{\it et al.}}
\def\jo{\journal}
\def\prl{Phys. Rev. Lett.}
\def\pr{Phys. Rev.}
\def\pl{Phys. Lett.}
\def\np{Nucl. Phys.}

\def\sj{Sov. J. Nucl. Phys.}

\def\apj{Astrophys. J.}

\def\yf{Yad. Fiz.}
\def\zp{Z. Phys. C}

\ref{M. S. Atiya \et, \jo\prl&66(91)2189.}
\ref{M. S. Atiya \et, \jo\prl&69(92)733.}
\ref{M. I. Dobrolyubov, \jo\yf&52(90)551 [\jo\sj&52(90)352]; M. I. Dobrolyubov
     and A. Yu. Ignatiev, \jo\pl&B206(88)346; \jo\zp&39(88)251;
     \jo\np&B309(88)655; A. E. Nelson and N. Tetradis, \jo\pl&B221(89)80.}
\ref{M. Suzuki, \jo\prl&56(86)1339; S. H. Aronson, H.-Y. Cheng, E. Fischbach,
     and W. Haxton, \jo\prl&56(86)1342; C. Bouchiat and J. Iliopoulos,
     \jo\pl&B169(86)447.}
\ref{B. Holdom, \jo\pl&B166(86)196.}
\ref{W. P. Lam and K.-W. Ng, \jo\pr&D44(91)3345.}
\ref{E. Fischbach, J. T. Gruenwald, S. P. Rosen, H. Spivack,
     A. Halprin, and B. Kayser, \jo\pr&D13(76)1523; \jo\pr&D16(77)2377.}
\ref{K. A. Olive, D. N. Schramm, G. Steigman, and T. P. Walker,
     \jo\pl&B236(90)454; T. P. Walker, G. Steigman, D. N. Schramm, K. A.
     Olive, and H. S. Kang, \jo\apj&51(91)376.}

\refout
\end